\documentclass{PoS}

\title{Exploring universality of transversity in p-p collisions}

\ShortTitle{Exploring universality of transversity in p-p collisions}

\author{\speaker{Marco Radici}%\thanks{A footnote may follow.}
\\
INFN Sezione di Pavia, via Bassi 6, I-27100 Pavia, Italy 
\\
E-mail: \email{marco.radici@pv.infn.it}}

\abstract{The transversity distribution was recently extracted from deep-inelastic scattering processes producing  hadron pairs in the final state. Together with a specific chiral-odd di-hadron fragmentation function, it is involved in the elementary mechanism that generates a transverse-spin asymmetry in the azimuthal distribution of the detected hadron pairs. The same elementary mechanism was predicted to generate an analogous asymmetry when the hadron pairs are produced in proton-proton collisions with one transversely polarized proton. Recently, the {\tt STAR} Collaboration has observed this asymmetry. We analyze the impact of these data on our knowledge of transversity.}

\FullConference{QCD Evolution 2016\\
		May 30-June 03, 2016\\
		National Institute for Subatomic Physics (Nikhef), Amsterdam}

\begin{document}

\section{Introduction}
\label{sec:intro}

Parton distribution functions (PDFs) describe combinations of number densities of quarks and gluons in a fast-moving hadron.  At leading twist, the quark structure of spin-$\textstyle{\frac{1}{2}}$ hadrons is described by three PDFs: the unpolarized distribution $f_1$, the longitudinal polarization (helicity) distribution $g_1$, and the transverse polarization (transversity) distribution $h_1$. From the phenomenological point of view, $h_1$ is the least known one because it is connected to QCD-suppressed processes where the parton helicity is flipped ({\it i.e.}, it is a chiral-odd function). Therefore, it can be measured only in processes with two hadrons in the initial state (e.g., proton-proton collision) or one hadron in the initial state and at least one hadron in the final state (e.g., semi-inclusive deep-inelastic scattering - SIDIS).  

Being chiral-odd, transversity has very peculiar features. There is no gluon transversity in a spin-$\textstyle{\frac{1}{2}}$ hadron like the nucleon. Hence, transversity scales with QCD evolution as a pure non-singlet function. The knowledge of its first Mellin moment, the so-called nucleon tensor charge, can influence current searches for physics beyond the Standard Model (BSM). In fact, high-precision low-energy measurements of $\beta-$decays or rare meson decays can be connected to BSM effects generated at TeV scales through effective Lagrangians that describe new semi-leptonic transitions, involving four-fermion contact terms or scalar/pseudo-scalar/tensor/$(V+A)$ interactions with operators up to dimension six (for a review, see Ref.~\cite{Cirigliano:2013}). The scalar and tensor operators contribute linearly to $\beta-$decay through their interference with SM operators, therefore they are more easily detectable. The transition amplitude is proportional to the product of the BSM coupling and of the corresponding hadronic charge; for the tensor interaction, we indicate them with $\epsilon_T$ and $g_T$, respectively. The experimental measurements have reached nowadays a per-mil precision level, and will perform better in the near future~\cite{Cirigliano:2013}. Therefore, it is important to determine the nucleon tensor charge $g_T$ with the largest possible precision in order to deduce reliable information on the unknown BSM coupling $\epsilon_T$~\cite{aurore}. 

Our phenomenological knowledge of transversity is rather limited. The Torino group extracted it for the first time by simultaneously fitting data on polarized single-hadron SIDIS and data on almost back-to-back emission of two hadrons in $e^+ e^-$ annihilations~\cite{Anselmino:2013}. The main difficulty of such analysis lies in the factorization framework used to interpret the data, since it involves Transverse Momentum Dependent PDFs (TMDs). QCD evolution of TMDs must be included to analyze SIDIS and $e^+ e^-$ data obtained at very different scales. But the computation is involved  and only recently an attempt to give a (not complete) description of these effects was released~\cite{Kang:2015msa}.

Alternatively, transversity can be extracted in the standard framework of collinear factorization using 
data on SIDIS with two hadrons detected in the final state. The cross section for di-hadron SIDIS at leading twist contains a term which is proportional to the product of $h_1$ and $H_1^{\sphericalangle}$, a specific chiral-odd Di-hadron Fragmentation Function (DiFF)~\cite{Jaffe:1998hf,Radici:2001na,Radici:2003lm}. The same $H_1^{\sphericalangle}$ appears in the leading-twist cross section for semi-inclusive back-to-back emission of hadrons pairs in $e^+ e^-$ annihilations~\cite{e+e-:2003}. The $H_1^{\sphericalangle}$ was parametrized from the $e^+ e^-$ data of {\tt BELLE} for production of $(\pi^+ \pi^-)$ pairs~\cite{noiBelle}. The valence components $h_1^{u_v}$ and $h_1^{d_v}$ of transversity were subsequently extracted from the {\tt HERMES} and {\tt COMPASS} data for SIDIS production of $(\pi^+ \pi^-)$ pairs~\cite{Bacchetta:2011ip,h1JHEP}. Recently, the analysis was updated by enclosing the latest and more precise {\tt COMPASS} data for a transversely polarized proton target~\cite{Radici:2015mwa}. 

In the context of di-hadron production, the extraction of transversity is based on the following assumption for the valence flavor $q_v$ at the starting scale $Q_0^2=1$ GeV$^2$: 
\begin{equation} 
x\, h_1^{q_v}(x; Q_0^2)=
\tanh \Bigl[ x^{1/2} \, \bigl( A_q+B_q\, x+ C_q\, x^2+D_q\, x^3\bigr)\Bigr]\, x \, 
\Bigl[ \mbox{\small SB}^q(x; Q_0^2)+ \mbox{\small SB}^{\bar q}(x; Q_0^2)\Bigr] \, .
\label{eq:funct_form}
\end{equation} 
The functional form fulfills the Soffer inequality at any scale (the explicit expression for the Soffer bound SB$^q$ can be found in the Appendix of Ref.~\cite{h1JHEP}). The low-$x$ behavior is determined by the $x^{1/2}$ term, which is imposed by hand to grant the integrability of Eq.~(\ref{eq:funct_form}). Present fixed-target data do not allow to constrain it. The extraction of $h_1$ was repeated using the functional form with one node ($A, B$ parameters, the socalled {\it rigid} scenario), two nodes ($A, B, C$ parameters, the {\it flexible} scenario), and three nodes ($A, B, C, D$ parameters, the {\it extra-flexible} scenario), and varying as well the normalization of the strong coupling constant at the $Z$ boson mass $(\alpha_S (M_Z^2))$ to account for the theoretical uncertainty in the determination of the $\Lambda_{\rm QCD}$ parameter. The outcomes turned out to be rather stable against these variations~\cite{Radici:2015mwa}. In the following, only the results with the {\it flexible} scenario will be shown. The error analysis was carried out using the replica method~\cite{h1JHEP,Radici:2015mwa}. 

%%%%%%%%%  Fig. 1  %%%%%%%%%
\begin{figure}[htb]
\begin{center}
\includegraphics[width=7cm]{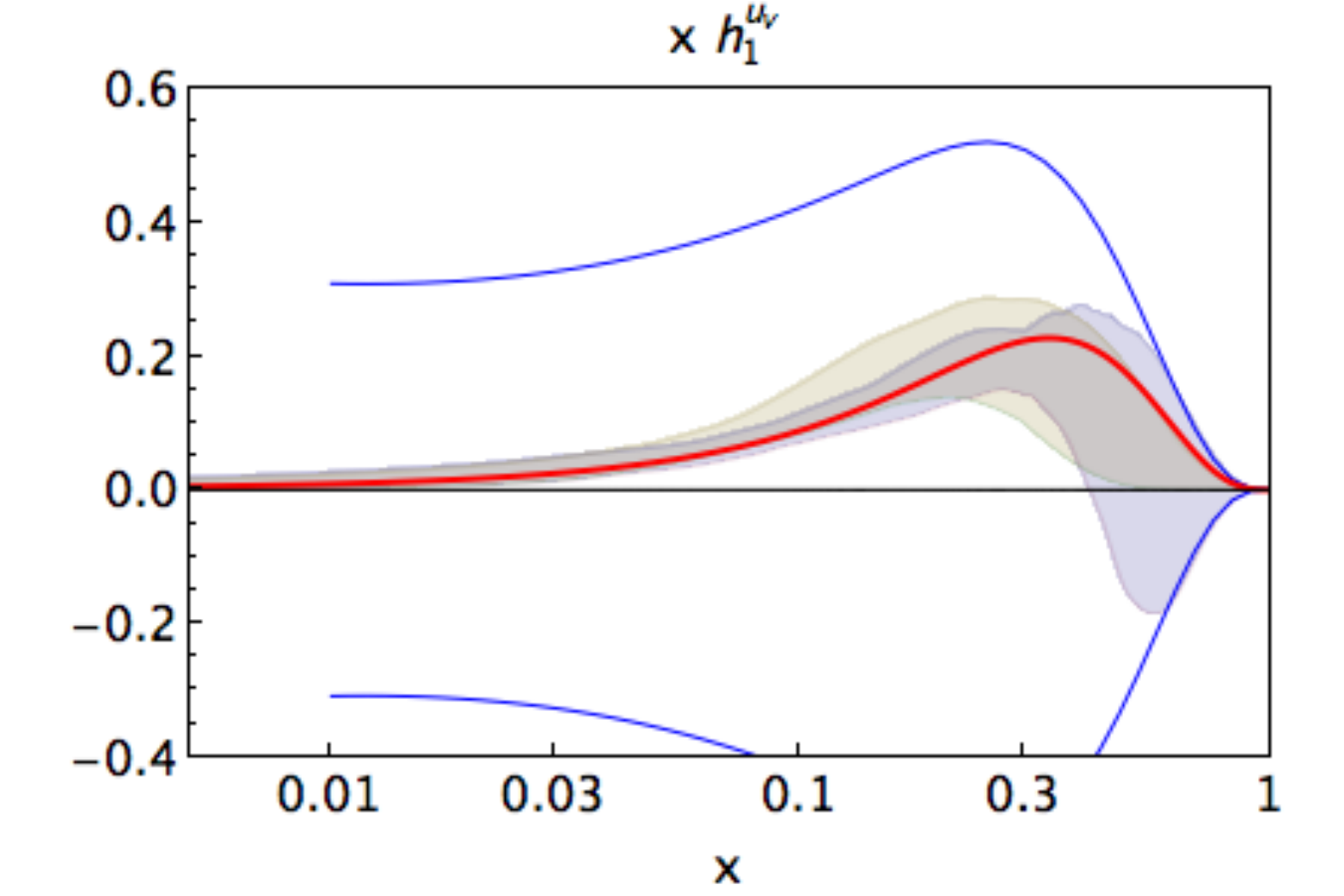}\hspace{1cm}\includegraphics[width=7cm]{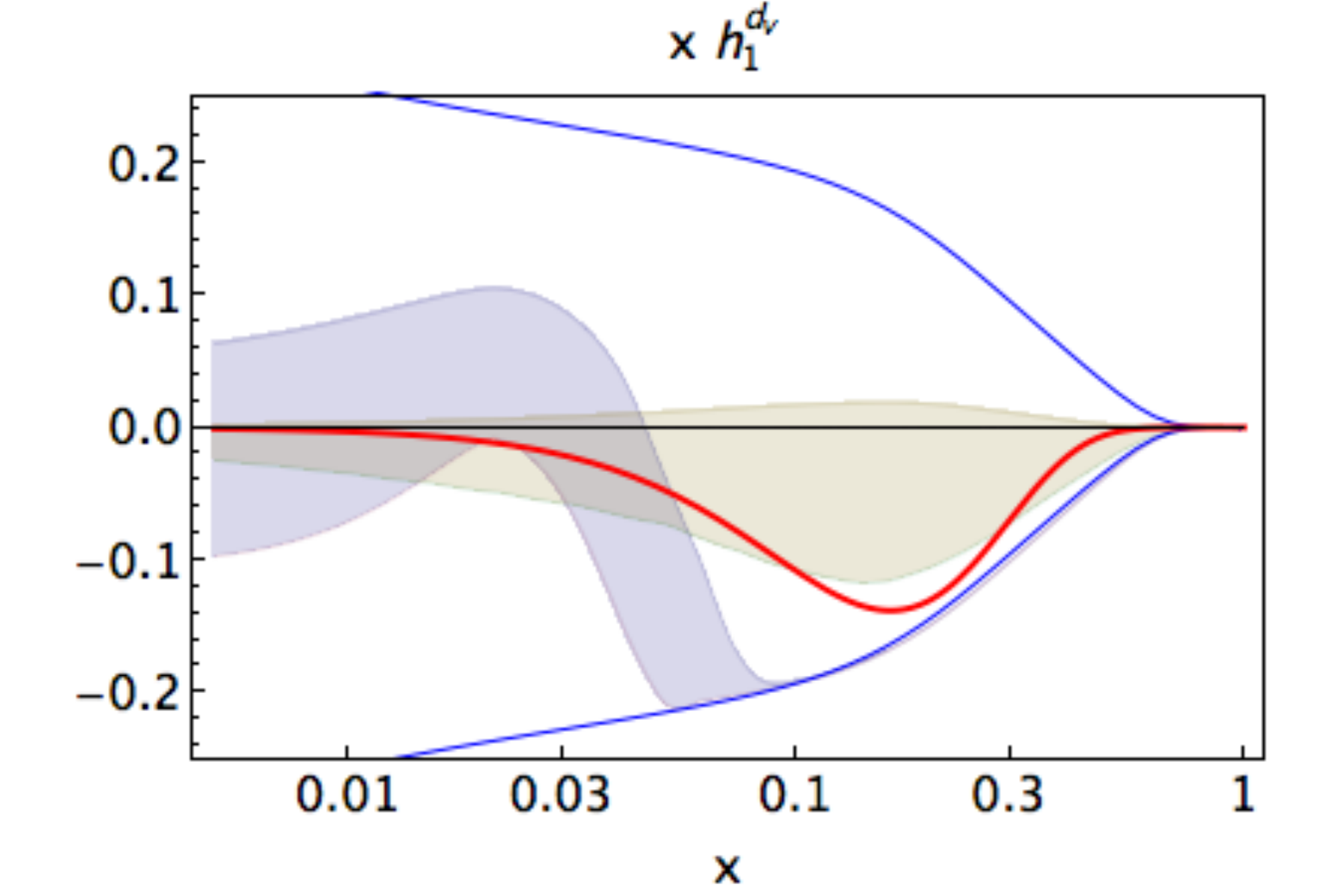}
\end{center}
\caption{Left panel: the $x h_1^{u_v} (x)$ at $Q^2=2.4$ GeV$^2$. The darker uncertainty band is the 68\% of replicas obtained in Ref.~\cite{Radici:2015mwa} with the {\it flexible} scenario and $\alpha_S (M_Z^2) = 0.139$~\cite{MSTW:2009}. The lighter band is the Torino extraction using the Collins effect~\cite{Anselmino:2013}. The central thick solid line is the result of Ref.~\cite{Kang:2015msa}. The dark thick solid lines indicate the Soffer bound. Right panel: same notations for the $d_v$ component.}
\label{fig:xh1}
\end{figure}
%%%%%

In Fig.~\ref{fig:xh1}, the left panel displays $x h_1^{u_v} (x)$ at $Q^2 = 2.4$ GeV$^2$. The darker uncertainty band is made of the central 68\% of all the replicas obtained in Ref.~\cite{Radici:2015mwa} with $\alpha_S (M_Z^2) = 0.139$~\cite{MSTW:2009}. The lighter band is the Torino extraction using the Collins effect~\cite{Anselmino:2013}. The central thick solid line is the result of Ref.~\cite{Kang:2015msa}, where evolution equations have been computed in the TMD framework (the related error band mostly overlaps with the lighter band). The dark thick solid lines indicate the Soffer bound. There is a general consistency among the various extractions, at least for $0.0065 \leq x \leq 0.34$ where the SIDIS data are. However, the error analysis based on the replica method~\cite{Radici:2015mwa} gives a more realistic description of the uncertainty, particularly outside the data range. It is worth to notice in the right panel that for $x \geq 0.1$ all replicas from the di-hadron extraction tend to saturate the lower limit of the Soffer bound. This trend is visible in all explored scenarios, indicating that it is not an artifact of the chosen functional form. Rather, it is driven by some of the fitted {\tt COMPASS} deuteron data, in particular by the bins n. 7 and 8, as we will show later in Sec.~\ref{sec:STAR}. 

The main advantage of extracting transversity based on the semi-inclusive di-hadron production lies in the possibility of working in a collinear factorization framework. The analysis can be extended also to hadronic collisions. Transversity can be studied in a process different from SIDIS and the universality of the elementary mechanism $h_1 \, H_1^{\sphericalangle}$ can be explored. This is a unique opportunity with respect to the extraction based on the Collins effect, because there are explicit counterexamples showing that TMD factorization is broken for single-hadron production in hadronic collisions~\cite{TMDnofac}.

%%%%

\section{Theoretical framework}
\label{sec:theory}

In the $p p^\uparrow \rightarrow (h_1\ h_2) X$ process, a proton with momentum $P_A$ collides on a transversely polarized proton with momentum $P_B$ and spin vector $S_B$, producing a pair of unpolarized hadrons $h_1, \, h_2,$ inside the same jet. The transverse component of the total pair momentum $P$ with respect to the beam $P_A$ is indicated with ${\bf P}_T$ and serves as the hard scale of the process. If the kinematics is collinear, namely if the transverse component of $P$ around the jet axis is integrated over, the differential cross section at leading order in $1/|{\bf P}_T|$ is~\cite{Bacchetta:2004it}
\begin{equation}
\frac{d\sigma}{d\eta\, d|{\bf P}_T|\, dM_h^2\,d\phi^{}_R\,d\phi^{}_{S_B}} =  
 d\sigma^0 \  \left( 1 + \sin (\phi^{}_{S_B} - \phi^{}_R) \, A_{pp} \right)  \; , 
\label{eq:ppcross}
\end{equation}
where $\eta$ is the pseudorapidity, $M_h^2 = P^2$ is the pair invariant mass, $\phi^{}_{S_B}$ is the azimuthal orientation of the spin vector $S_B$, and $\phi^{}_R$ describes the azimuthal orientation around ${\bf P}$ of the plane containing the hadron pair momenta (see Fig.~1 of Ref.~\cite{Bacchetta:2004it}). The unpolarized cross section $d\sigma^0$ is given by
\begin{equation}
\frac{d\sigma^0}{d\eta\, d|{\bf P}_T|\, dM_h^2} =  2 \, |{\bf P}_T| \, \sum_{a,b,c,d}\int \frac{d x_a\, dx_b }{4 \pi^2 z_h} \, f_1^a (x_a) \, f_1^b(x_b) \, \frac{d\hat{\sigma}_{ab \to cd}}{d\hat{t}} \, D_1^c (z_h, M_h^2) 
\label{eq:ppcross0}
\end{equation}
and the transverse spin asymmetry $A_{pp}$ by 
\begin{equation}
A_{pp}= \frac{2 \, |{\bf P}_T|\, |{\bf S}_{BT}|}{d\sigma^0} \, \frac{|{\bf R}_T|}{M_h}\, \sum_{a,b,c,d}\, \int \frac{dx_a \, dx_b}{16 \pi z_h} \, f_1^a(x_a) \, h_1^b(x_b) \, \frac{d\Delta \hat{\sigma}_{ab^\uparrow \to c^\uparrow d}}{d\hat{t}} H_1^{\sphericalangle c}(z_h, M_h^2) \, .
\label{eq:App}
\end{equation}
The ${\bf R}_T$ is the transverse component of the relative momentum of the hadron pair. Its modulus is a function of the invariant mass $M_h$~\cite{Radici:2003lm,Bacchetta:2004it}. The elementary cross section $d\hat{\sigma}$ describes the annihilation of partons $a$ and $b$ (carrying fractional momenta $x_a$ and $x_b$, respectively) into partons $c$ and $d$. The unpolarized DiFF $D_1^c$ describes the inclusive fragmentation of parton $c$ into the detected hadron pair. The $d\Delta\hat{\sigma}$ and $H_1^{\sphericalangle\, c}$ are the corresponding elements of the scattering amplitude when one of the partons is transversely polarized. All possible combinations of flavors $a + b \rightarrow c + d$ must be summed up (see Ref.~\cite{Bacchetta:2004it} for the analytic expression of the corresponding (polarized) elementary cross sections). Finally, the fractional energy $z_h$ carried by the hadron pair is constrained by momentum conservation in the elementary annihilation as~\cite{Bacchetta:2004it}
\begin{equation}
z_h = \frac{|{\bf P}_T|}{\sqrt{s}}\, \frac{x_a e^{-\eta} + x_b e^{\eta}}{x_a x_b} \, , 
\label{eq:zh}
\end{equation}
where $s = (P_A + P_B)^2$ is the squared cm energy of the collision, and $\hat{t} = t \  x_a / z_h$ with $t = (P_A - P_B)^2$.

%%%%

\section{Universality of transversity}
\label{sec:results}

Experimental evidence for the predicted transverse spin asymmetry $A_{pp}$ of Eq.~(\ref{eq:App}) has been recently reported by the {\tt STAR} collaboration for the process $p + p^\uparrow \to (\pi^+ \pi^-) + X$ at the cm energy $\sqrt{s} = 200$ GeV~\cite{Adamczyk:2015hri}. The predictions for $A_{pp}$ are obtained by replacing the $h_1^b, \, D_1^c, \, H_1^{\sphericalangle\, c}$ in Eqs.~(\ref{eq:ppcross0}, \ref{eq:App}) with those ones used to fit the {\tt BELLE} data for the production of $(\pi^+ \pi^-)$ pairs in $e^+ e^-$ annihilations~\cite{noiBelle}, and the {\tt HERMES} and {\tt COMPASS} data for the corresponding SIDIS case~\cite{Radici:2015mwa}. In the following pictures, each uncertainty band is formed by the central 68\% of all replicas of DiFFs and transversity that fit the above $e^+ e^-$ and SIDIS data, after their parametric expression at $Q_0^2 = 1$ GeV$^2$ is evolved to each $|{\bf P}_T|$ scale of {\tt STAR} data using standard DGLAP evolution~\cite{Ceccopieri:2007ip}.

%%%%%%%%%% Fig. 2  %%%%%%%%%
\begin{figure}[htb]
\begin{center}
\includegraphics[width=7cm]{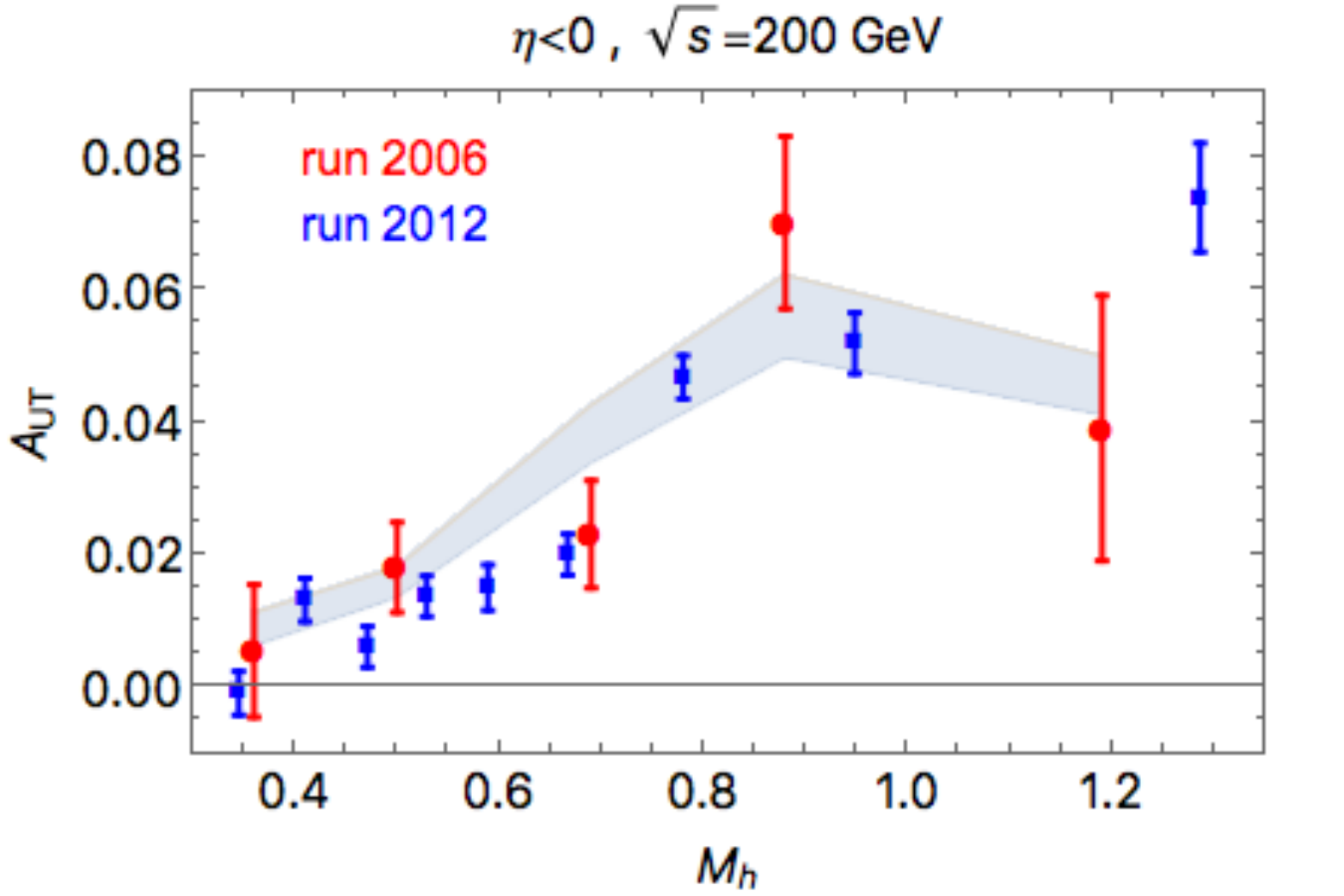}\hspace{1cm}\includegraphics[width=7cm]{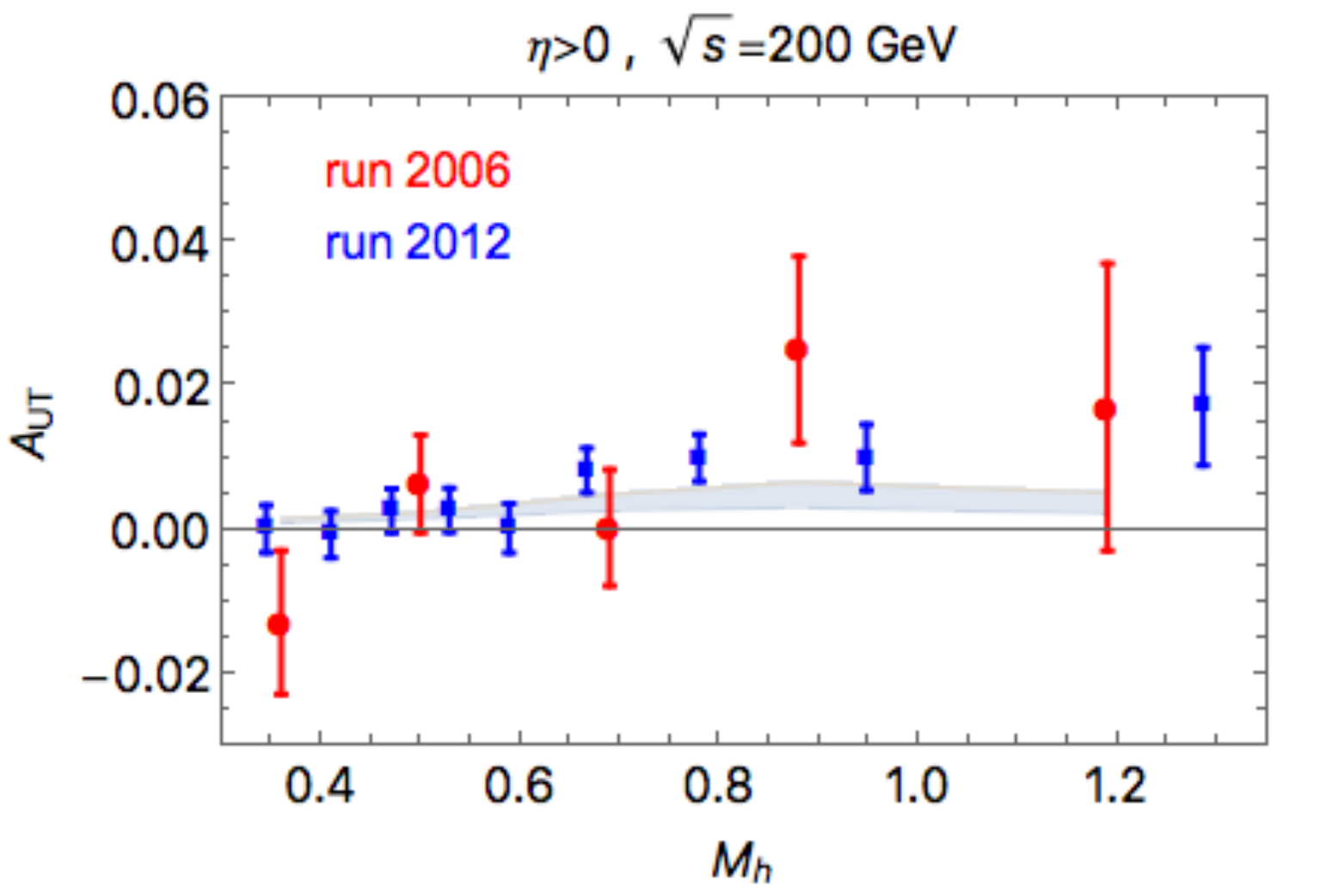}
\end{center}
\caption{The spin asymmetry $A_{pp}$ of Eq.~(2.3) as function of $M_h$ with $|{\bf P}_T|$ and $\eta$ integrated in each experimental bin. Left panel for negative (forward) $\eta$, right panel for positive (backward) $\eta$. Uncertainty band from 68\% of all replicas that fit $e^+ e^-$ and SIDIS data for $(\pi^+ \pi^-)$ production~[18]. Data points obtained from {\tt STAR} at $\sqrt{s} = 200$ GeV: the ones with "run 2006" label from Ref.~[16], the ones with "run 2012" label are preliminary results~[19].}
%\caption{The spin asymmetry $A_{pp}$ of Eq.~(\ref{eq:App}) as function of $M_h$ with $|{\bf P}_T|$ and $\eta$ integrated in each experimental bin. Left panel for negative (forward) $\eta$, right panel for positive (backward) $\eta$. Uncertainty band from 68\% of all replicas that fit $e^+ e^-$ and SIDIS data for $(\pi^+ \pi^-)$ production~\cite{noipp}. Data points obtained from {\tt STAR} at $\sqrt{s} = 200$ GeV: the ones with "run 2006" label from Ref.~\cite{Adamczyk:2015hri}, the ones with "run 2012" label are preliminary results~\cite{STARrun12}.}
\label{fig:AppMh}
\end{figure}
%%%%%

In Fig.~\ref{fig:AppMh}, the transverse spin asymmetry $A_{pp}$ of Eq.~(\ref{eq:App}) is plotted as function of the invariant mass $M_h$, integrating on $|{\bf P}_T|$ and on $\eta$. For each experimental $M_h$ bin, the theoretical result is obtained by integrating on $M_h$ over the bin width. The left panel refers to negative $\eta$, the right panel only to positive $\eta$. Positive pseudorapidities correspond to backward-propagating transversely polarized particles: the asymmetry is dominated by low$-x_b$ partons where the transversity is small and the resulting asymmetry is largely suppressed. At $\eta < 0$, the asymmetry is larger because it is dominated by the valence components of transversity. The experimental data have been collected by the {\tt STAR} collaboration for collisions at cm energy $\sqrt{s} = 200$ GeV. The data points with "run 2006" label have been published in Ref.~\cite{Adamczyk:2015hri}. Those ones with "run 2012" label are preliminary~\cite{STARrun12}. The  band shows the theoretical prediction for $A_{pp}(M_h)$~\cite{noipp}: it agrees in sign and shape with the experimental data. At $\eta < 0$, it shows the typical increase around the mass of the $\rho$ resonance. The overall agreement is found also at backward kinematics ($\eta > 0$) when looking at $A_{pp}$ as a function of $|{\bf P}_T|$~\cite{noipp}. It confirms the prediction in Ref.~\cite{Bacchetta:2004it}, namely that the mechanism producing the $A_{pp}$ asymmetry observed in the {\tt STAR} data is the same one that is responsible for the (spin) azimuthal asymmetries measured by the {\tt HERMES}, {\tt COMPASS}, and {\tt BELLE} collaborations in SIDIS and $e^+ e^-$ processes, respectively.

%%%%%%%%%% Fig. 3  %%%%%%%%%
\begin{figure}[htb]
\begin{center}
\includegraphics[width=7cm]{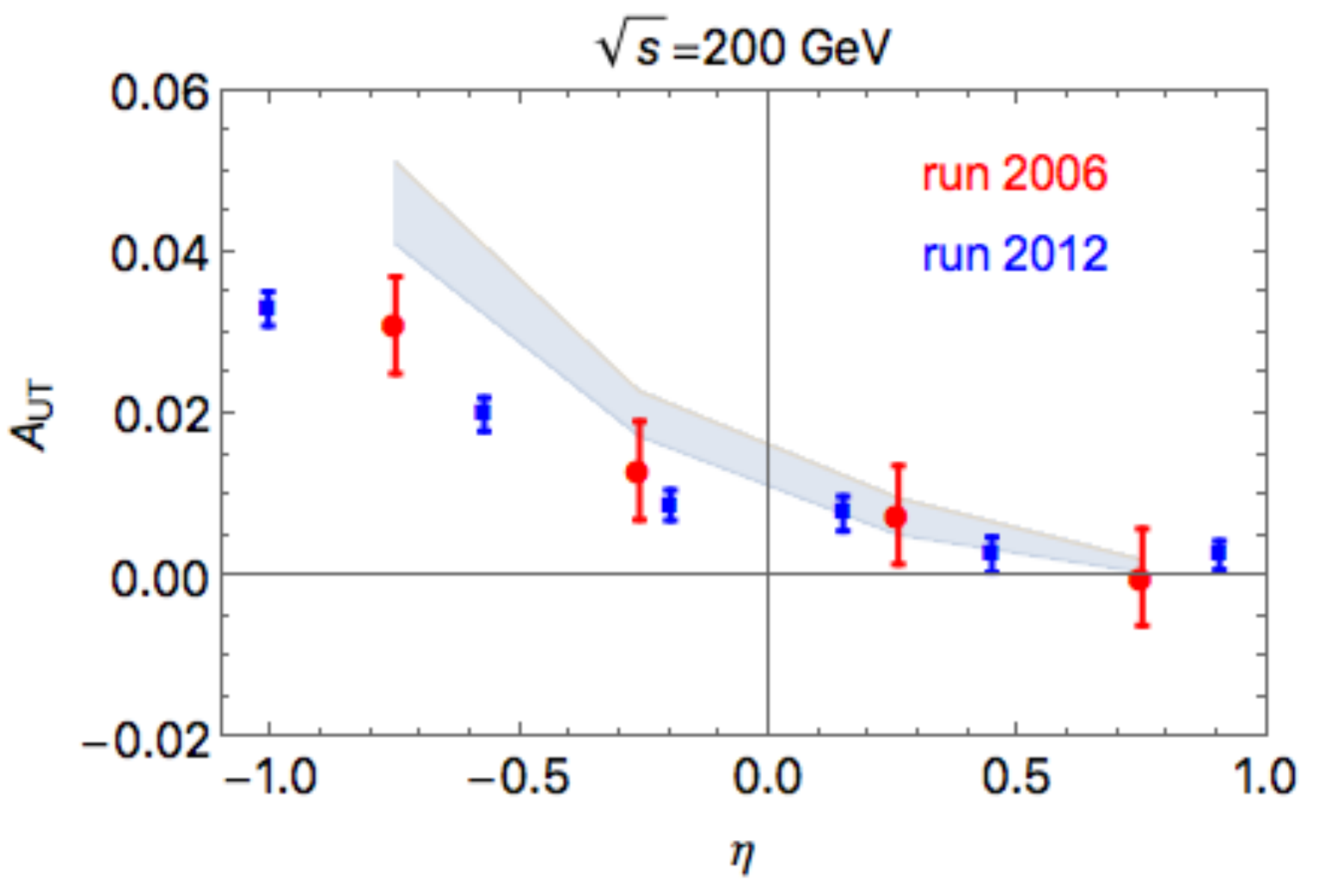}\hspace{1cm}\includegraphics[width=7cm]{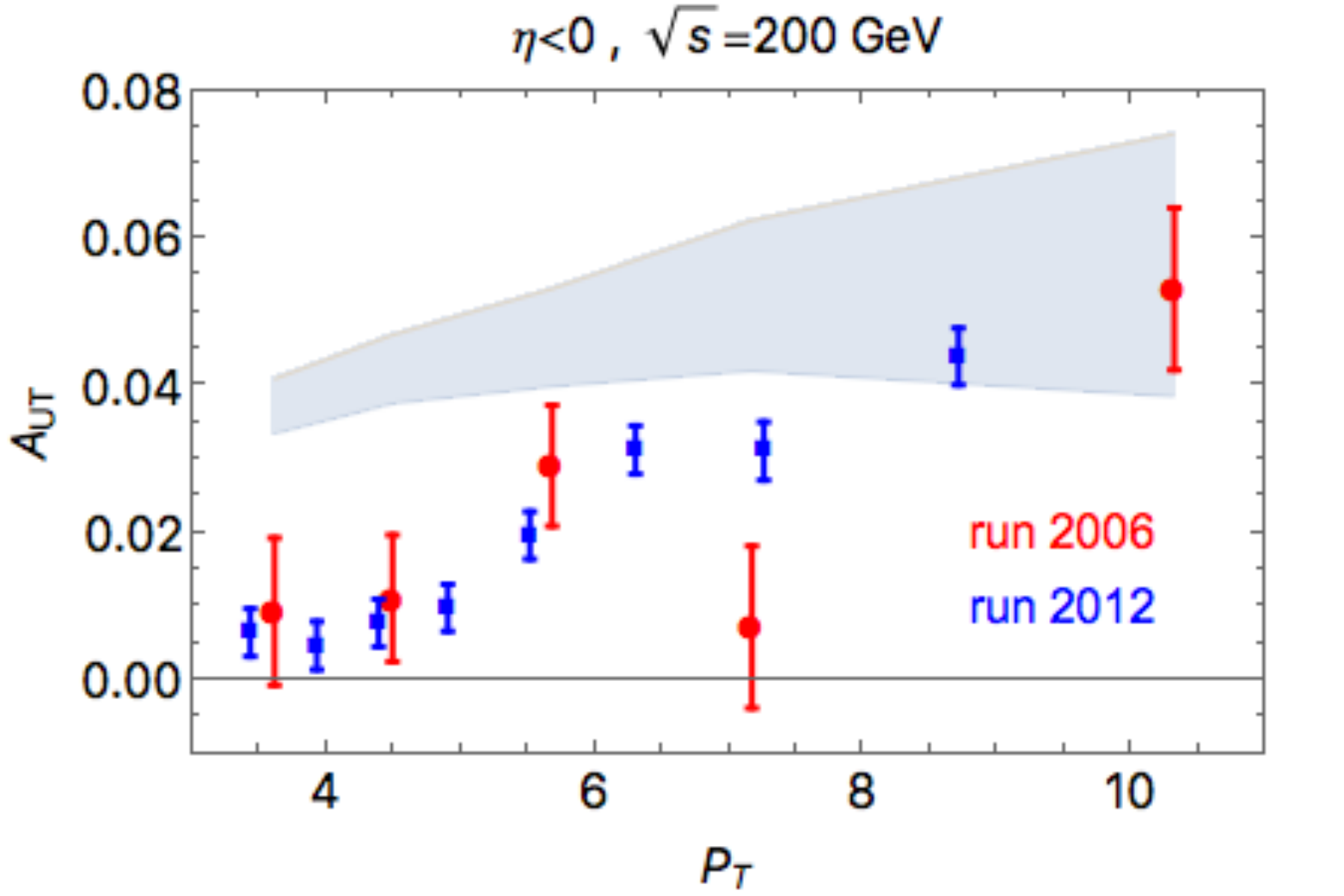}
\end{center}
\caption{Left panel: the spin asymmetry $A_{pp}$ of Eq.~(2.3) as function of $\eta$ with $|{\bf P}_T|$ and $M_h$ integrated in each experimental bin. Forward kinematics corresponds to negative $\eta$. Right panel: $A_{pp}$ as function of $|{\bf P}_T|$ with $M_h$ and only negative $\eta$ integrated in each experimental bin. Notation and conventions as in previous figure.}
%\caption{Left panel: the spin asymmetry $A_{pp}$ of Eq.~(\ref{eq:App}) as function of $\eta$ with $|{\bf P}_T|$ and $M_h$ integrated in each experimental bin. Forward kinematics corresponds to negative $\eta$. Right panel: $A_{pp}$ as function of $|{\bf P}_T|$ with $M_h$ and only negative $\eta$ integrated in each experimental bin. Notation and conventions as in previous figure.}
\label{fig:Appeta}
\end{figure}
%%%%%

In Fig.~\ref{fig:Appeta}, the left panel shows the asymmetry $A_{pp}$ of Eq.~(\ref{eq:App}) as a function of $\eta$ when integrating on ${\bf P}_T|$ and $M_h$. Similarly to the previous figure, for each experimental $\eta$ point the theoretical result is integrated upon the corresponding bin. At $\eta < 0$, the agreement is less satisfactory: the 68\% band of computed replicas starts to deviate from the experimental points. One encounters the same problem when considering the predictions for the ${\bf P}_T|$ dependence of the asymmetry in the same forward kinematics $(\eta < 0)$, as shown in the right panel. These discrepancies could be explained in terms of the currently missing higher-order QCD corrections to $d\sigma^0$, which are usually expected to be large for hadronic collisions. These corrections should not affect the numerator of $A_{pp}$ because there is no gluon transversity, so no compensation should occur when taking the ratio to form the asymmetry. Moreover, $D_1^g$, the gluon channel in di-hadron fragmentation, is poorly constrained: no data of unpolarized cross sections for the semi-inclusive $(\pi^+ \pi^-)$ production are available either in $e^+ e^-$ annihilations or in $p-p$ collisions. The $D_1^g$ contribution is currently determined only through QCD evolution~\cite{noipp}. 

This lack of information points out that the theoretical uncertainty band could be larger than the indicated one, paradoxically improving the compatibility with data. However, it is important to remark that no $K$ factor could ever modify the agreement displayed in Fig.~\ref{fig:AppMh} about the dependence of the asymmetry upon the pair invariant mass. Therefore, the crucial message about the universality of the $h_1 \, H_1^{\sphericalangle}$ mechanism is still valid~\cite{noipp}.

%%%%

\section{The impact of RHIC data}
\label{sec:STAR}

In this section, we reconsider the problem outlined when commenting Fig.~\ref{fig:xh1}. Namely, the fact that the valence down component of each one of the 68\% of replicas obtained from fitting the {\tt HERMES} and {\tt COMPASS} SIDIS data saturates the Soffer bound already at the scale $Q^2 = 2.4$ GeV$^2$ and inside the $x-$range of SIDIS data. We can verify that this unusual trend is induced by two specific bins (the n. 7 and 8) in the set of {\tt COMPASS} data collected with a transversely polarized deuteron target.

%%%%%%%%%% Fig. 4  %%%%%%%%%
\begin{figure}[htb]
\begin{center}
\includegraphics[width=7cm]{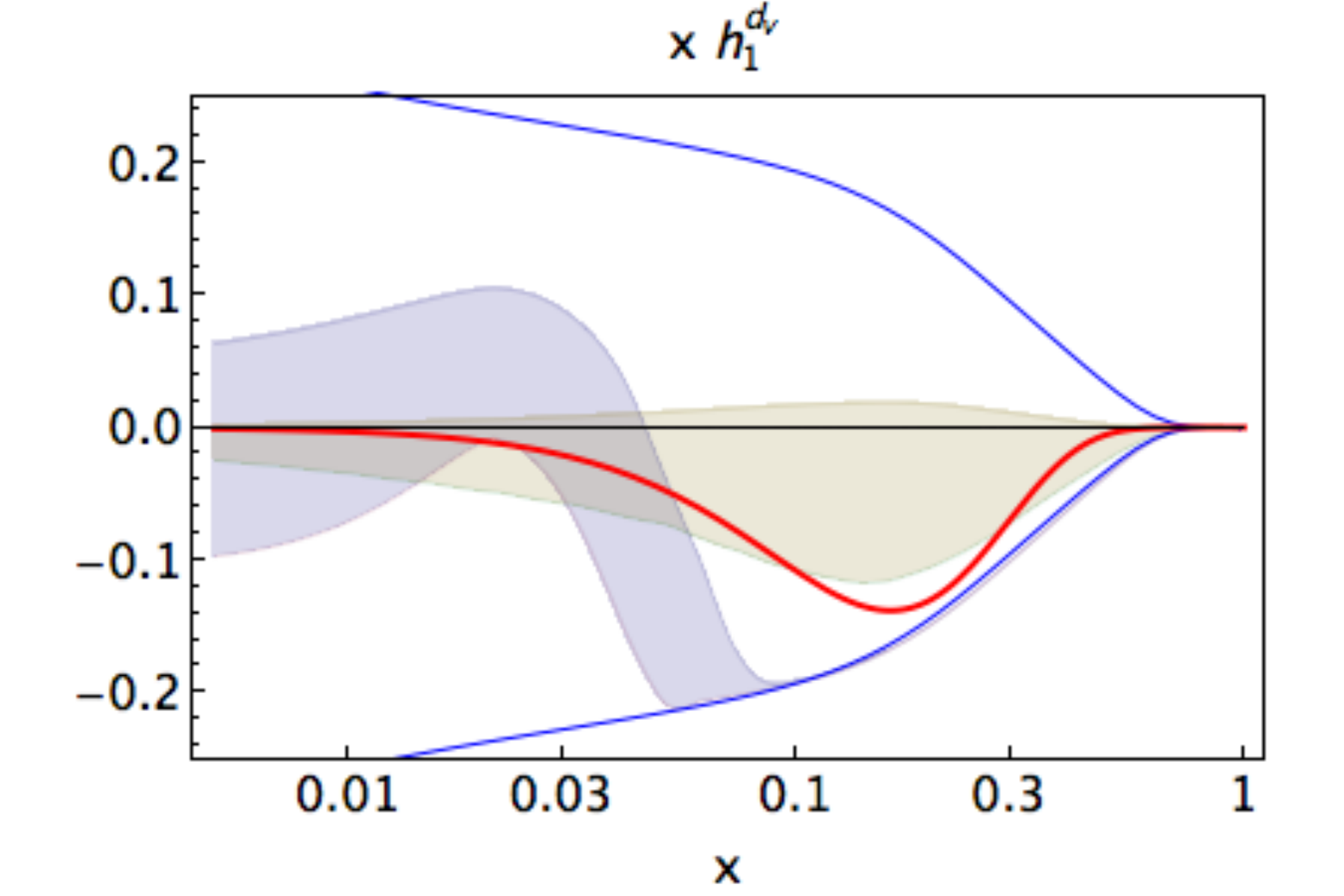}\hspace{1cm}\includegraphics[width=7cm]{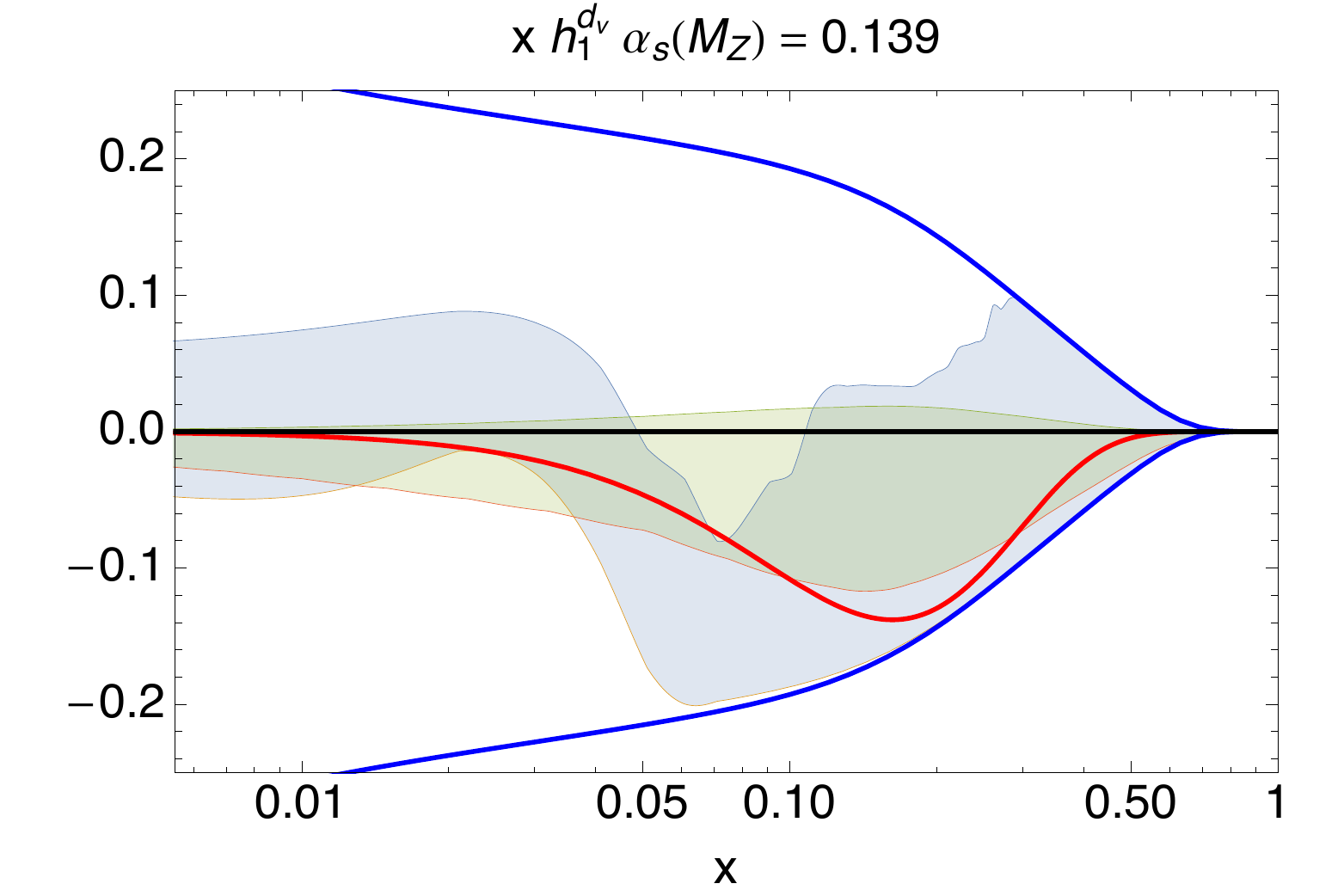}
\end{center}
\caption{Left panel: the $x h_1^{d_v} (x)$ at $Q^2=2.4$ GeV$^2$ with the same notation and conventions as in Fig.~1. Right panel: the same quantity as it is extracted from the fit when the bins n. 7,8 from {\tt COMPASS} data for deuteron target are excluded.}
%\caption{Left panel: the $x h_1^{d_v} (x)$ at $Q^2=2.4$ GeV$^2$ with the same notation and conventions as in Fig.~\ref{fig:xh1}. Right panel: the same quantity as it is extracted from the fit when the bins n. 7,8 from {\tt COMPASS} data for deuteron target are excluded.}
\label{fig:xh1no78}
\end{figure}
%%%%%

In Fig.~\ref{fig:xh1no78}, the left panel is identical to the right panel of Fig.~\ref{fig:xh1}. Namely, the darker band shows the valence down component of the central 68\% of all replicas obtained by fitting the {\tt HERMES} and {\tt COMPASS} SIDIS data for inclusive $(\pi^+ \pi^-)$ pair production. The right panel shows what we would get by excluding from the fit the bins n. 7, 8 of the {\tt COMPASS} deuteron data around $Q^2 \sim 10$ GeV$^2$. The replicas would not prematurely saturate the Soffer bound, on the contrary they would fluctuate within the lower and upper bounds showing a better degree of compatibility with the transversity extraction based on the Collins effect.

%%%%%%%%%% Fig. 5  %%%%%%%%%
\begin{figure}[htb]
\begin{center}
\includegraphics[width=7cm]{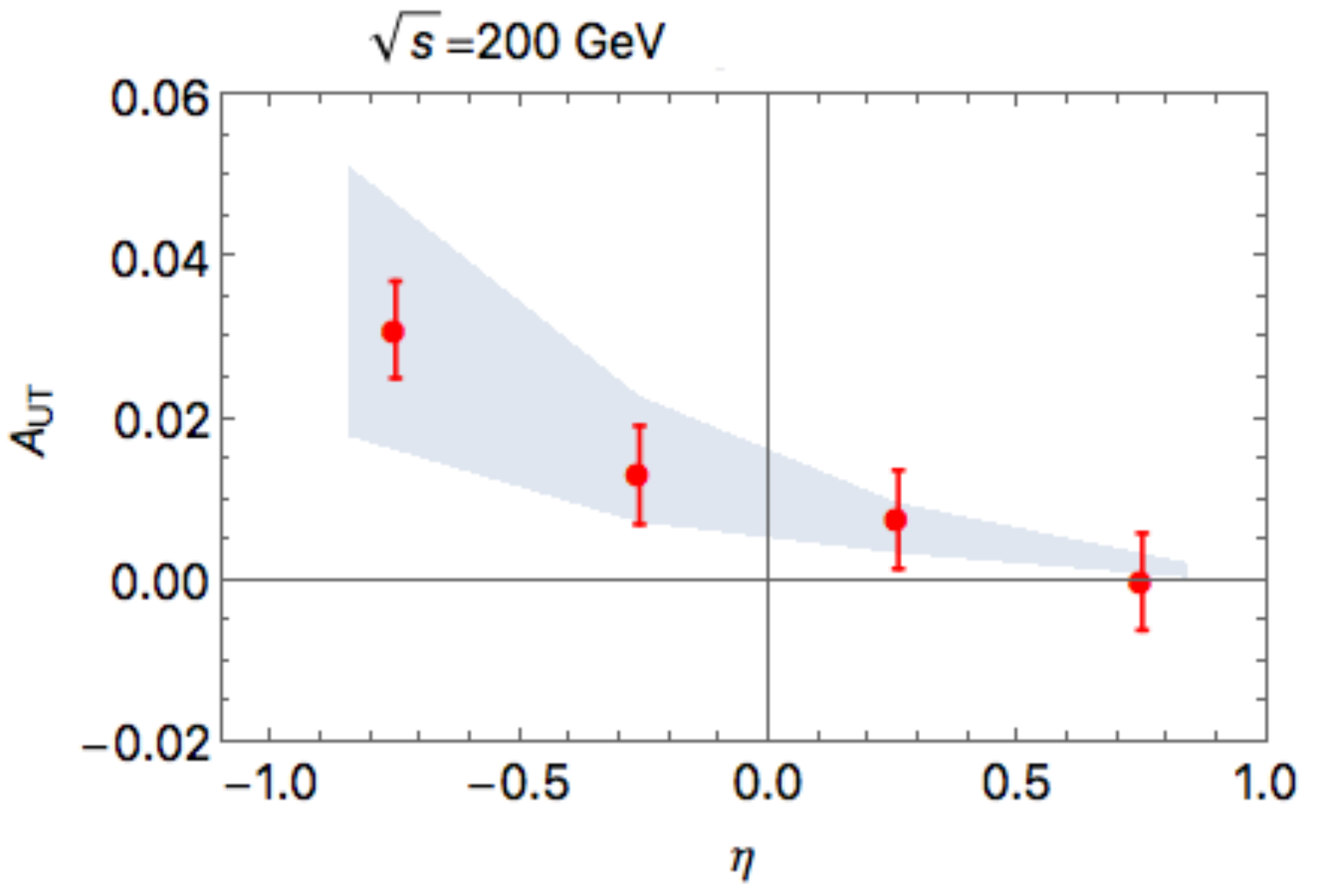}\hspace{1cm}\includegraphics[width=7cm]{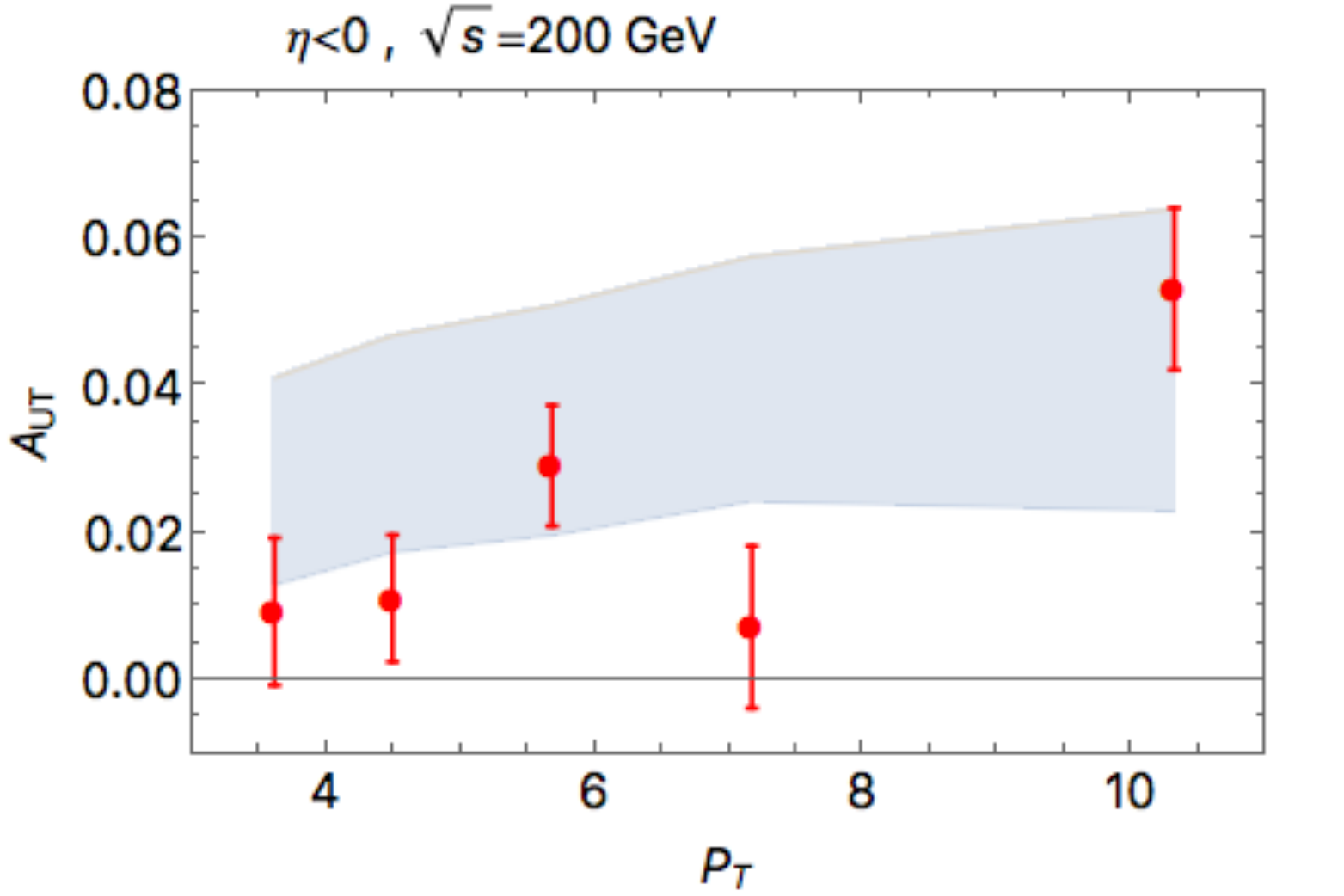}
\end{center}
\caption{Left panel: the spin asymmetry $A_{pp}$ of Eq.~(2.3) as function of $\eta$ with $|{\bf P}_T|$ and $M_h$ integrated in each experimental bin. Right panel: $A_{pp}$ as a function of $|{\bf P}_T|$ with $M_h$ and only negative $\eta$ integrated in each experimental bin. Notations and conventions as in Fig.~3. The uncertainty band corresponds to the central 68\% of all replicas fitting the $e^+ e^-$ and SIDIS data when the bins n. 7, 8 of {\tt COMPASS} deuteron data are excluded.}
%\caption{Left panel: the spin asymmetry $A_{pp}$ of Eq.~(\ref{eq:App}) as function of $\eta$ with $|{\bf P}_T|$ and $M_h$ integrated in each experimental bin. Right panel: $A_{pp}$ as a function of $|{\bf P}_T|$ with $M_h$ and only negative $\eta$ integrated in each experimental bin. Notations and conventions as in Fig.~\ref{fig:Appeta}. The uncertainty band corresponds to the central 68\% of all replicas fitting the $e^+ e^-$ and SIDIS data when the bins n. 7, 8 of {\tt COMPASS} deuteron data are excluded.}
\label{fig:Appetano78}
\end{figure}
%%%%%

The message of Fig.~\ref{fig:xh1no78} can be simply reformulated by saying that if bins n. 7, 8 of {\tt COMPASS} deuteron data are excluded from the fit, the resulting replicas show a higher degree of flexibility. This feature not only increases the level of compatibility with other extractions of transversity based on the Collins effect, but also has an impact on the compatibility with {\tt STAR} data for proton-proton collisions. In Fig.~\ref{fig:Appetano78}, the same situation of Fig.~\ref{fig:Appeta} is reconsidered but the uncertainty band of theoretical predictions is now made of the central 68\% of all replicas obtained when the bins n. 7, 8 of the {\tt COMPASS} deuteron data set are excluded from the fit. More flexibility corresponds to a better compatibility also with the {\tt STAR} data. This finding should not be misinterpreted as if a certain subset of {\tt COMPASS} SIDIS data should be discarded. Rather, it suggests that the {\tt STAR} data can be strongly selective on the results of the fit to the SIDIS data themselves. A further evidence of this can be found by re-examining the results shown in Fig.~\ref{fig:Appeta}.

%%%%%%%%%% Fig. 6  %%%%%%%%%
\begin{figure}[htb]
\begin{center}
\includegraphics[width=7cm]{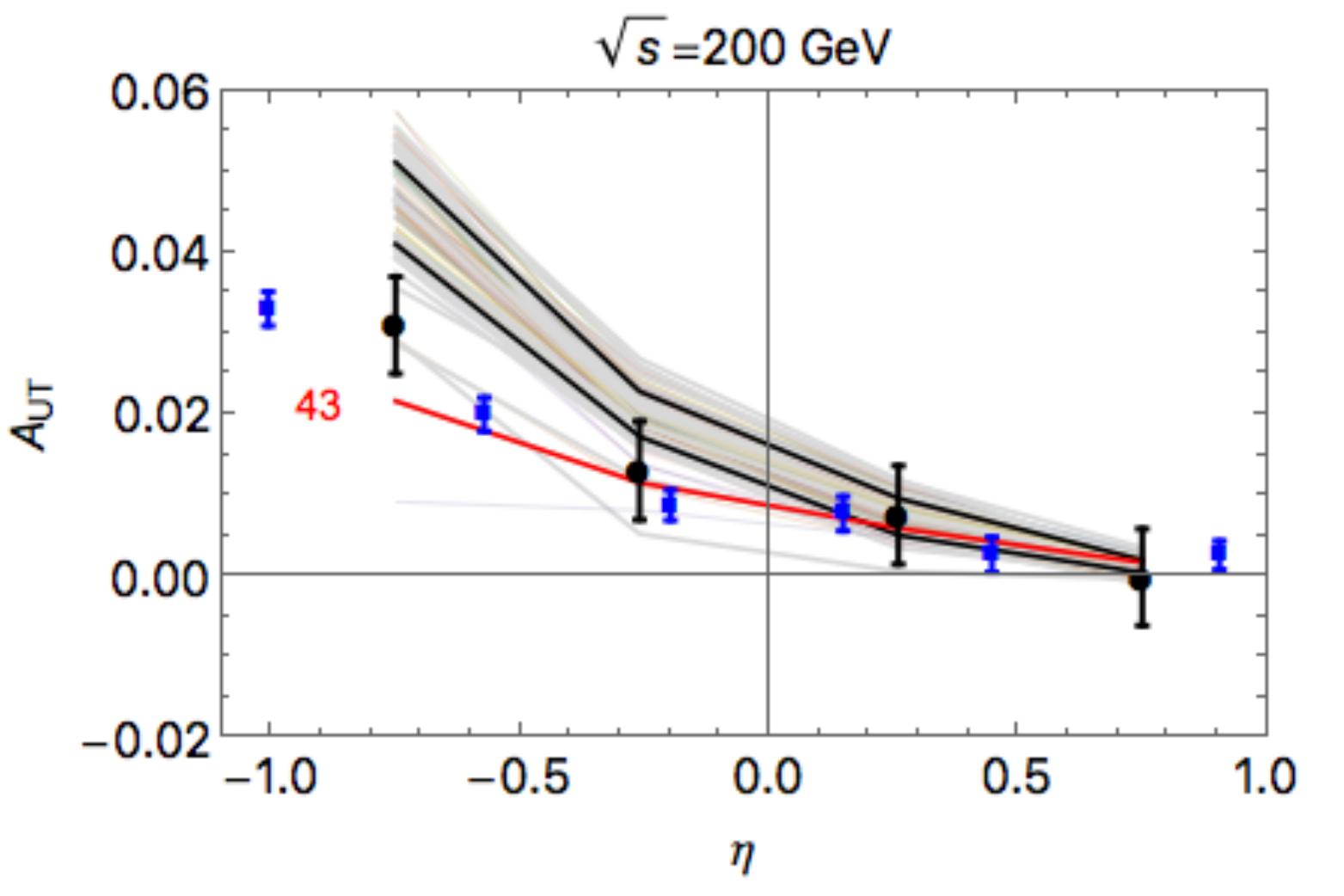}\hspace{1cm}\includegraphics[width=7cm]{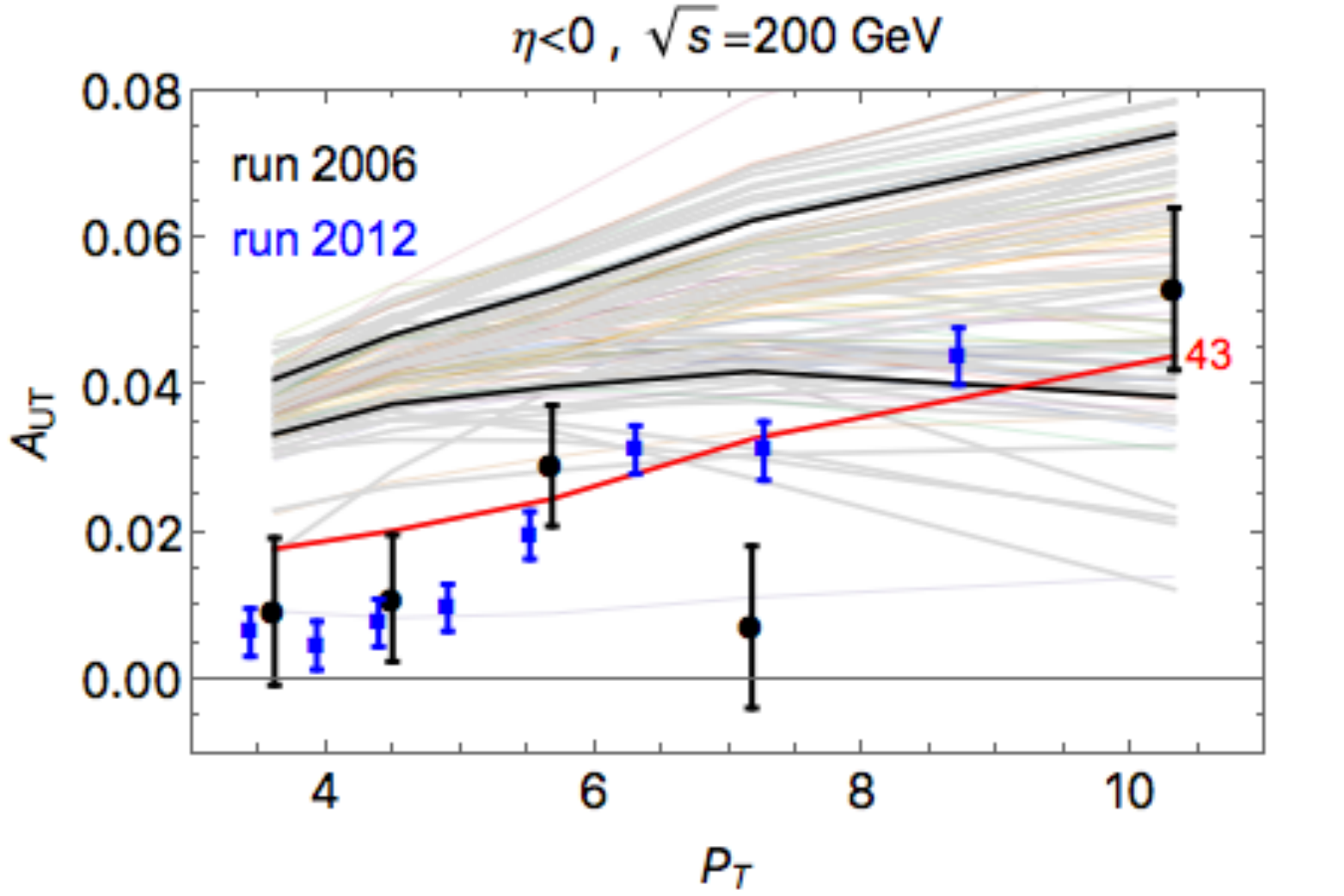}
\end{center}
\caption{Same plot as in Fig.~3, but now all replicas fitting the $e^+ e^-$ and SIDIS data are drawn, including also those ones that lie outside the central 68\% band represented by the two thick solid lines. One of this replicas is marked by the label "43".}
%\caption{Same plot as in Fig.~\ref{fig:Appeta}, but now all replicas fitting the $e^+ e^-$ and SIDIS data are drawn, including also those ones that lie outside the central 68\% band represented by the two thick solid lines. One of this replicas is marked by the label "43".}
\label{fig:Appetaall}
\end{figure}
%%%%%

Fig.~\ref{fig:Appetaall} shows the same plot of Fig.~\ref{fig:Appeta} but now all replicas fitting the $e^+ e^-$ and SIDIS data are drawn, including also those ones that lie outside the central 68\% band represented by the two thick solid lines. Among them, we note that the one marked with the label "43" is compatible with almost all the {\tt STAR} data measured at $\sqrt{s} = 200$ GeV. Indeed, this means that data from $p-p$ collisions can bring a lot of additional information to what we already know about transversity from the fit of $e^+ e^-$ and SIDIS data. 

This argument can be turned into a quantitative statement by using the technique of reweighting the replicas~\cite{NNPDF}. In short, each replica obtained from the fit to $e^+ e^-$ and SIDIS data is assigned a probability weight according to its $\chi^2$ on the {\tt STAR} data. The probability weight measures the statistical relevance of each considered replica. Some of them will have very low weights and will be discarded. This loss of efficiency is measured through the so-called Shannon entropy. Moreover, the distribution of computed $\chi^2$ of replicas must be peaked around unity in order for the {\tt STAR} data to be compatible with the $e^+ e^-$ and SIDIS ones. 

%%%%%%%%%% Fig. 7  %%%%%%%%%
\begin{figure}[htb]
\begin{center}
\includegraphics[width=6cm]{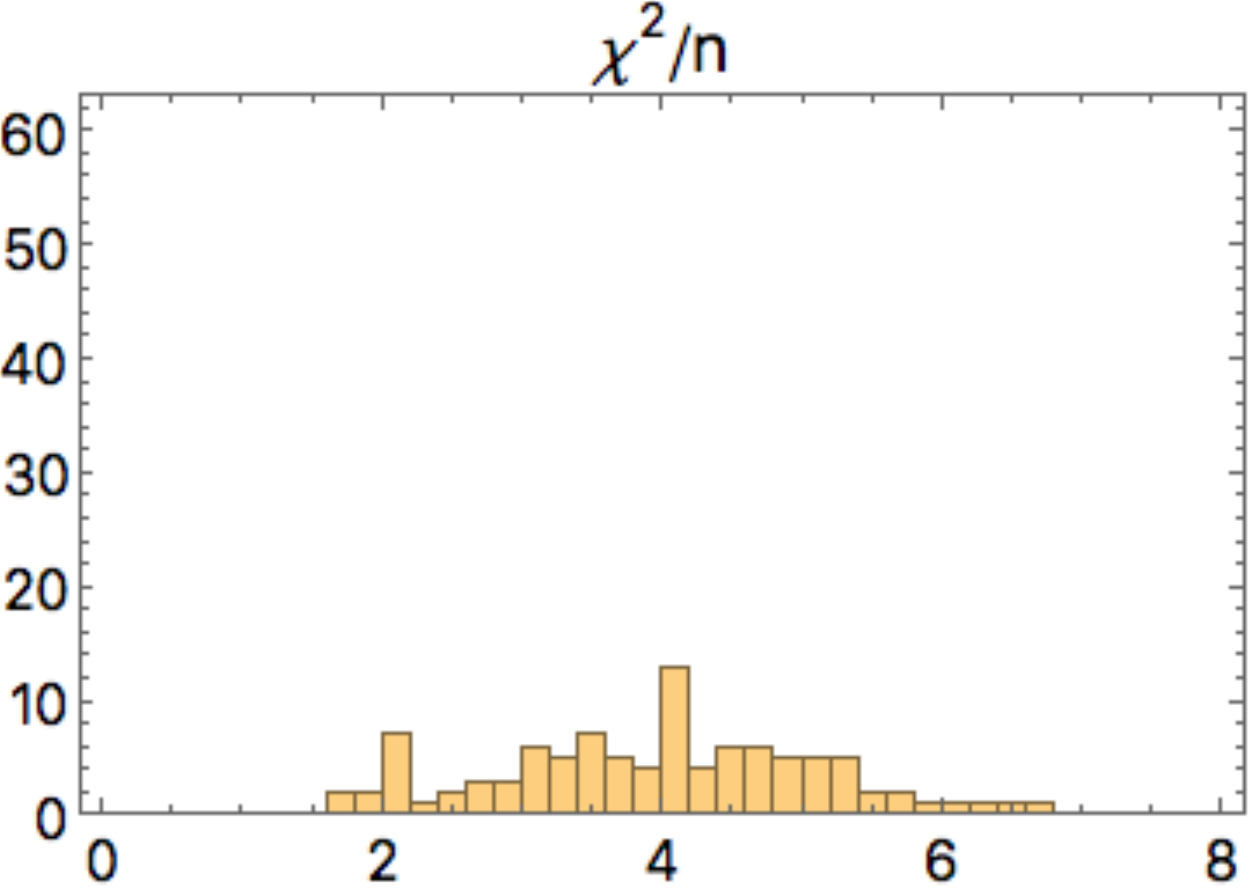}\hspace{1cm}\includegraphics[width=6cm]{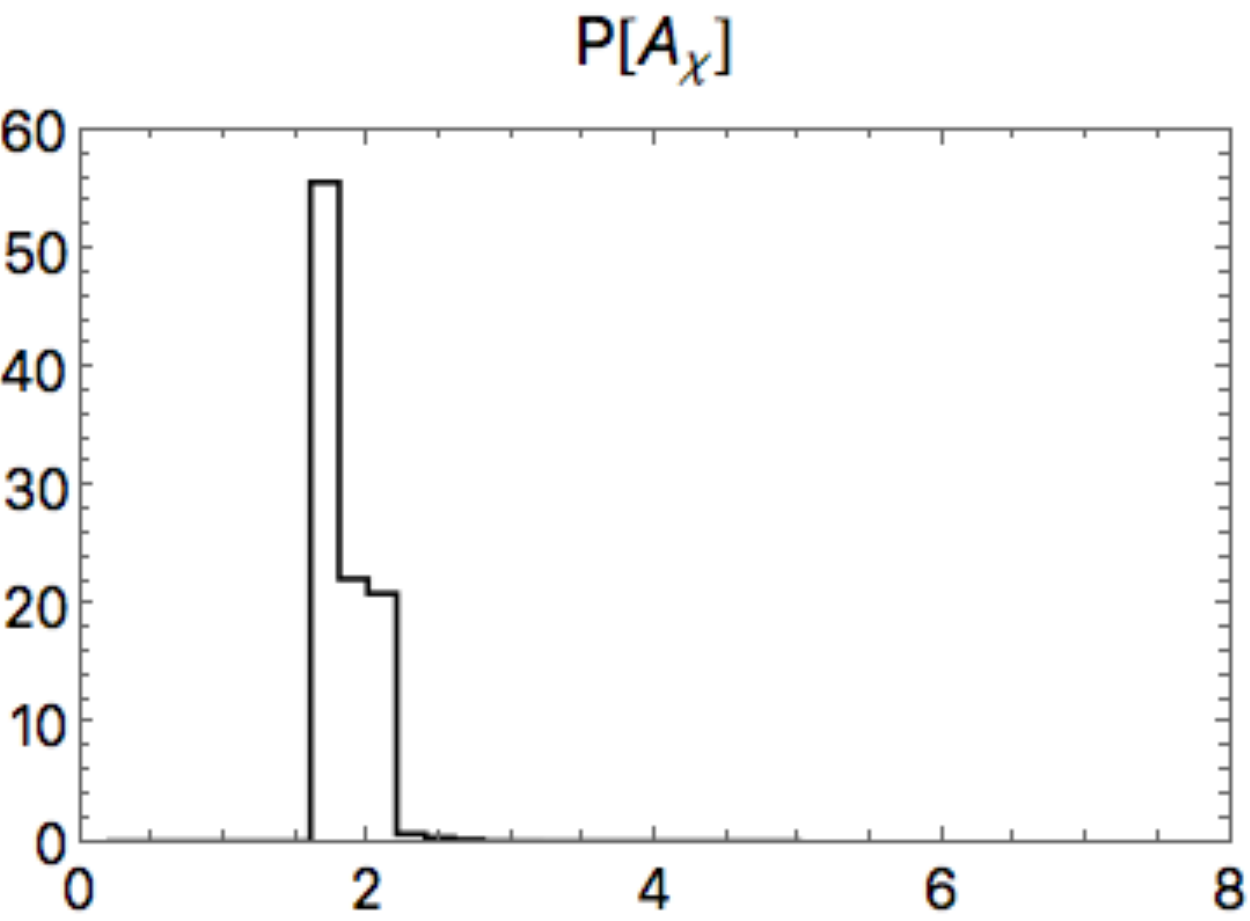}
\end{center}
\caption{Left panel: the normalized distribution of $\chi^2$ on $n$ {\tt STAR} data~\cite{Adamczyk:2015hri} of all replicas computed by fitting the $e^+ e^-$ and SIDIS data for di-hadron production~\cite{Radici:2015mwa}. Right panel: the distribution of $\chi^2$ after reweighting the replicas.}
\label{fig:reweight}
\end{figure}
%%%%%

In Fig.~\ref{fig:reweight}, the left panel shows the distribution of $\chi^2$ computed on the {\tt STAR} data (and normalized to the number of data points) for each replica derived by fitting the data for $(\pi^+ \pi^-)$ pair production from $e^+ e^-$ annihilations ({\tt BELLE}) and SIDIS ({\tt HERMES} and {\tt COMPASS} with deuteron bins n. 7,8 excluded). The right panel shows how the distribution changes after reweighting the set of replicas. The peak is centered slightly below 2, suggesting a reasonable degree of consistency among the considered data sets. But the computed Shannon entropy indicates that only 7\% of the replicas is statistically relevant. It means that the information brought by the {\tt STAR} data statistically selects only few of the replicas extracted by fitting the $e^+ e^-$ and SIDIS data.

%%%%%

\section{Conclusions}
\label{sec:end}

The transversity distribution can be reliably extracted by fitting the combined set of data for the semi-inclusive production of di-hadron pairs in $e^+ e^-$ annihilations and SIDIS. The extraction is performed by working in the collinear factorization framework. It can be used as a cross-check of the extraction performed in the TMD factorization framework and based on the Collins effect. But it uniquely allows to extend the analysis of di-hadron production also to proton-proton collision data. 

Based on a suitable factorization formula, a specific spin asymmetry was predicted for the azimuthal distribution of di-hadron pairs produced from proton-proton collisions where one of the two protons is transversely polarized. This asymmetry has been recently measured by the {\tt STAR} collaboration for the production of $(\pi^+ \pi^-)$ pairs. Using the results of the fit to $e^+ e^-$ and SIDIS data, predictions for the asymmetry in proton-proton collisions show a high degree of compatibility with the {\tt STAR} measurements. It means that the same universal elementary mechanism is active in all hard processes leading to the production of $(\pi^+ \pi^-)$ pairs. 

Using the reweighting technique, it is possible to "measure" which is the impact of the proton-proton collision data on the amount of knowledge accumulated by fitting the $e^+ e^-$ and SIDIS data. Anyway, it is desirable to perform a global fit of all hard processes involving transversity and inclusively producing di-hadron pairs. Since the theoretical framework is based on collinear factorization, this goal is technically achievable, and work is in progress along this line.

%%%%

\section*{Acknowledgments}

Most of the results presented in this report have been carried out in collaboration with A.~ Bacchetta, A.~Mukherjee, and A.M. Ricci, to whom I am deeply indebted. This research is partially supported by the European Research Council (ERC) under the European Union's Horizon 2020 research and innovation program (Grant Agreement No. 647981, 3DSPIN).

%%%%

\end{document}